\documentclass{iopart}
\usepackage[T1]{fontenc}
\usepackage{graphicx}
\usepackage{url,graphics,epsfig}
\usepackage{amssymb}

\usepackage{tikz}

\newcommand*\mycirc[1]{%
  \begin{tikzpicture}[baseline=(C.base)]
    \node[draw,circle,inner sep=1pt](C) {#1};
  \end{tikzpicture}}

\begin{document}

\jl{2}
%
%
%
\def\etal{{\it et al~}}
%
%
%
%
%
%
\setlength{\arraycolsep}{2.5pt}             

\title[{K-Shell Photoionization of  Singly Ionized Atomic Nitrogen}]{K-Shell Photoionization 
	of  Singly Ionized Atomic Nitrogen: Experiment and Theory}

\author{M F Gharaibeh$^1\footnote[1]{Corresponding author, E-mail: mfg@just.edu.jo}$, 
              J M Bizau$^{2,3}$,  D Cubaynes$^{2,3}$,  S Guilbaud$^{2}$,  N El Hassan$^{2}$,  
              M M Al Shorman$^{2}$, C Miron$^{3}$, C Nicolas$^{3}$, E Robert$^{3}$,
              C Blancard$^{4}$ and B M McLaughlin$^{5,6}\footnote[2]{Corresponding author, E-mail: b.mclaughlin@qub.ac.uk}$}

\address{$^1$Department of Physics, Jordan University of Science and Technology, Irbid 22110, Jordan.}

\address{$^2$Institut des Sciences Mol\'{e}culaires d'Orsay (ISMO), CNRS UMR 8214, 
			Universit\'{e} Paris-Sud, B\^{a}t. 350, F-91405 Orsay cedex, France}

\address{$^{3}$Synchrotron SOLEIL - L'Orme des Merisiers, Saint-Aubin - BP 48 91192 Gif-sur-Yvette cedex, France}

\address{$^{4}$CEA-DAM-DIF, Bruy$\grave{\rm e}$res-le-Ch\^{a}tel, F-91297 Arpajon Cedex, France}

\address{$^{5}$Centre for Theoretical Atomic, Molecular and Optical Physics (CTAMOP),\\
			School of Mathematics and Physics, Queen's University Belfast, 
			Belfast BT7 1NN, Northern Ireland, UK}

\address{$^{6}$Institute for Theoretical Atomic and Molecular Physics (ITAMP),\\
			Harvard Smithsonian Center for Astrophysics, MS-14, Cambridge, MA 02138, USA}


%
%

\begin{abstract}
Absolute cross sections for the K-shell photoionization of C-like nitrogen ions were measured by
employing the ion-photon merged-beam technique at the
SOLEIL synchrotron radiation facility in Saint-Aubin, France.
High-resolution spectroscopy with E/$\Delta$E $\approx$ 7,000 was achieved with
the photon energy from 388 to 430 eV scanned with a band pass of 300 meV, 
and the 399.4 to 402 eV range with 60 meV.
 Experimental results are compared  with theoretical predictions 
made from the multi-configuration Dirac-Fock (MCDF) and R-matrix methods. 
The interplay between experiment  and theory enabled the identification 
and characterization of the strong 1s  $\rightarrow$ 2p resonances observed 
in the spectra.  
\end{abstract}

%
%



\vspace{1.0cm}
\begin{flushleft}
Short title: K-shell photoionization of N$^{+}$ ions\\
\vspace{1cm} Draft for J. Phys. B: At. Mol. \& Opt. Phys: \today
\end{flushleft}

\maketitle
%
%
%
\section{Introduction}
Satellites {\it Chandra} and {\it XMM-Newton} currently provide a wealth of X-ray spectra of 
astronomical objects; however, a serious lack of high-quality atomic data impedes
the interpretation of these spectra
 \cite{McLaughlin2001, Brickhouse2010, Kallman2010, Kallman2011, Quinet2011}. 
Recent studies on Carbon and its ions have shown that high quality data 
is required to model the observations in the X-ray spectrum of the bright blazar
Mkn 421 observed by the Chandra LETG+HRC-S \cite{McLaughlin2010}.
Spectroscopy in the soft X-ray region (5-45 \AA) including 
K-shell transitions of C, N, O, Ne, S and Si, in neutral, singly or doubly ionized states and L-shell 
transitions of Fe and Ni, provides a valuable tool for probing the extreme 
environments in active galactic nuclei (AGN's),
X-ray binary systems, cataclysmic variable stars (CV's) and Wolf-Rayet Stars \cite{Skinner2010}
as well as the interstellar media (ISM) \cite{Garcia2011}. The latter, recent work, for example, 
demonstrated that X-ray spectra from {\it XMM-Newton} can be used to characterize ISM, 
provided accurate atomic oxygen {\it K}-edge cross sections are available. 
Analogous results concerning the chemical composition of the ISM are to be expected 
with the availability of accurate data on neutral \cite{Witthoeft2009,McLaughlin2011} 
and singly ionized atomic nitrogen {\it K}-edge cross sections. 
The globular cluster X-ray source CXO J033831.8-352604 in NGC 1399 has also recently
been found to show strong emission lines of [O III] and [N II] in its optical 
spectrum in addition to ultraluminous X-ray emission with a soft X-ray spectrum \cite{Irwin2010,Warner2011}.

This lack of  available X-ray data has motivated not only the present authors to perform K-shell 
photoionization studies on such ions but also
other groups who use the same or similar experimental techniques. 
K-shell photoionization cross section results have been obtained on a variety of ions of astrophysical interest;
He-like Li$^{+}$ \cite{Scully2006,Scully2007},
Li-like  B$^{2+}$ \cite{Mueller2010}, C$^{3+}$  \cite{Mueller2009}, 
Be-like B$^{+}$ \cite{Mueller2011} , C$^{2+}$ \cite{Scully2005},
B-like C$^{+}$ \cite{Schlachter2004},
N-like O$^{+}$ \cite{Kawatsura2002},
F-like Ne$^{+}$ \cite{Yamaoka2001},
along with valence shell studies on Mg-like Fe$^{14+}$ \cite{Simon2010}. 
The majority of this experimental data has been benchmarked with 
state-of-the-art theoretical methods.

The present  work on this proto-type C-like ion provides benchmark 
values for cross sections on photoabsorption of X-rays in 
the vicinity of the {\it K}-edge, where strong n=2 inner-shell resonance states of singly ionized atomic nitrogen are observed. No experiments have been reported to date on singly ionized atomic nitrogen 
in the {\it K}-edge photon energy region, previous experimental studies have been 
restricted to the valence shell region \cite{Kjeldsen2002}.

Promotion of  a K-shell electron in C-like nitrogen (N$^\mathrm{+}$) ions
 to an outer np-valence shell (1s $\rightarrow$ np)  from the ground
state produces states that can autoionize, forming a N$^\mathrm{2+}$ ion and an outgoing free electron.
One of the strongest excitation processes in the interaction of a photon
with the $\rm 1s^22s^22p^2~^3P$ ground-state of the C-like nitrogen
ion is the 1s $\rightarrow$ 2p photo-excitation process;
$$
 h\nu + {\rm N^{+}(1s^22s^22p^2~^3P)}  \rightarrow  {\rm N^{+} ~ (1s2s^2 \,2p^3 ~^3S^o, ^3P^o, ^3D^o) }
 $$
 $$
 \downarrow
 $$
 $$
{\rm  N^{2+}~ (1s^22s^22p~^2P^o) + e^-.}
$$
Experimental studies of this singly ionized atomic nitrogen ion in its ground state
$\rm 1s^22s^22p^2~^3P$  are further hampered by the presence of 
metastable states  as the N$^+$ ions are produced 
in the gas-phase using an Electron-Cyclotron-Resonance-Ion-Source (ECRIS). 
Metastable states $\rm 1s^22s^22p^2~^1D, \,^1S$,  and $\rm 1s^22s2p^3~^5S^o$   are present 
in the ion beam.  In the case of the $\rm 1s^22s2p^3~^5S^o$ metastable 
state auto-ionization processes occuring by the 1s $\rightarrow$ 2p photo-excitation process  are;
$$
 h\nu + {\rm N^{+}(1s^22s2p^3~^5S^o)}
 $$
$$
\downarrow
$$
$$
{\rm  N^{+} [1s2s[^{1,3}S]\,2p^4(^3P,^1D,^1S)]~^{5}P}
$$
$$
\downarrow
$$
$$
{\rm N^{2+} (1s^22s2p^2~ ^4P) + e^-.}
$$
Similarly,  auto-ionization processes may also occur for the $\rm 1s^22s^22p^2~ ^1D, \, ^1S$ 
metastable states interacting with photons.  To our knowledge this would appear to be  
the first time experimental measurements have been performed 
on this proto-type C-like system in the photon energy region of the {\it K}-edge.

Theoretical photoionization (PI) cross section calculations for inner-shell processes in this C-like ion 
have be performed by Reilman and Manson \cite{Reilman1979}
using the Hartree-Slater wavefunctions of Herman and Skillman \cite{HS1963} and
by Verner and co-workers \cite{Verner1993} using Dirac-Slater potentials 
(\cite{Slater1960,Band1979}) within a central field approximation.
Photoionization cross sections obtained using these methods may be suitably accurate 
at very high photon energies but often give poor results near thresholds where configuration 
mixing is strong and resonance structure prevails, as is the case for the processes investigated here
 in the vicinity of the K-edge threshold.  
Central field methods neglect resonances features (observed in experimental measurements) as degenerate subshells 
of equivalent electrons rather than individual energy levels are considered. 
Therefore, any results obtained from these approaches in spectral 
modelling should be treated with due caution.  

In this paper we present detailed measurements of the single and double PI cross sections 
in the 398 - 406 eV region (the only one where peaks were observed)
from  the entire 388 - 430 eV photon energy range explored.  
Our MCDF and R-matrix calculations enabled the identification and characterization 
of the strong $\rm 1s \rightarrow 2p$ resonances 
observed in the spectra. The present investigation provides absolute values (experimental and theoretical)
for PI cross sections,  n=2 inner-shell resonance energies and linewidths occurring in the interaction of a photon with the
$\rm 1s^22s^22p^2~^3P\,, ^1D\,,^1S$,  and $\rm 1s^22s2p^3~^5S^o$ states of the N$^{+}$ ion.

The layout of this paper is as follows. Section 2 details the experimental procedure used. 
Section 3 presents a brief outline of the theoretical work. Section 4 presents a discussion of the
results obtained from both the experimental and theoretical methods.
Finally in section 5 conclusions are drawn from the present investigation.

\section{Experiment}\label{sec:exp}
\subsection{Ion production}
\begin{figure}
\begin{center}
\includegraphics[width=10.0cm,angle=+90]{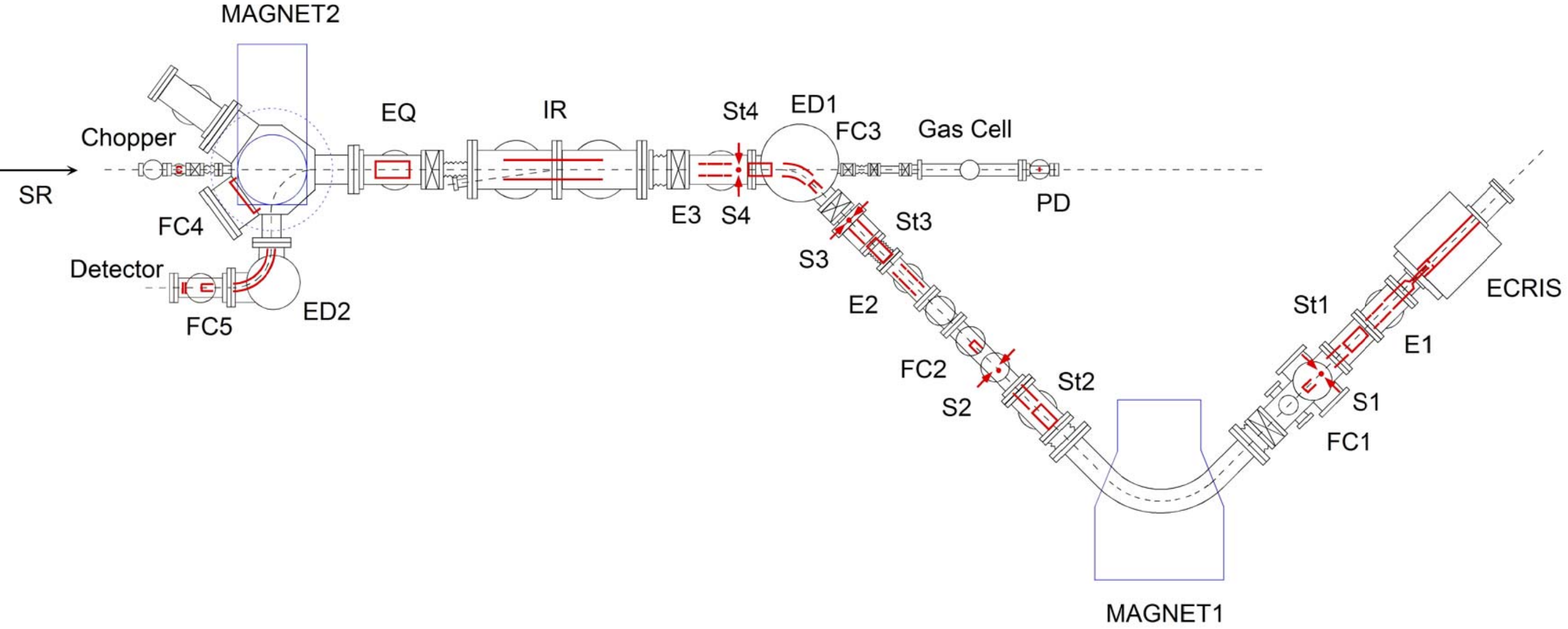}
\caption{\label{fig:setup}(Colour online) Scheme of the MAIA set-up on the PLEIADES beam line. 
								The ionic optics used for the ion beam transport are represented in red. 
								ECRIS: Electron Cyclotron Ion Source, E: einzel lens, 
								St: set of horizontal and vertical steerers, 
								S: collimating slits, FC: Faraday cup, 
								ED: electrostatic deflector, IR: interaction region, 
								EQ: electrostatic quadrupole, 
								SR: synchrotron radiation, PD: photodiode.\label{figure_setup}}
\end{center}
\end{figure}

The present measurements were made using the new MAIA (Multi-Analysis Ion Apparatus) 
set-up, figure \ref{figure_setup}, permanently installed on 
Branch A of the PLEIADES beam line \cite{PLEIADES} 
at SOLEIL, the French national synchrotron radiation facility, located in Saint-Aubin,
France.  It is a merged beam set-up, similar to the one originally designed by Peart {\it et al} \cite{Peart1973} for 
measuring electron impact cross sections. A simplified scheme of the set-up is shown in figure~\ref{fig:setup}.

The N$^{+}$ ions are produced in a permanent magnet Electron Cyclotron Resonance Ion Source (ECRIS) especially 
designed for this set-up at the CEA(Commissariat \'{a} l'Energie Atomique et aux Energies Alternatives) in Grenoble and previously tested at LURE 
(Laboratoire pour l'Utilisation de Rayonnement Electromagnétique) in Orsay \cite{BizauPRA2006}. 
To produce a nitrogen plasma, molecular nitrogen gas is injected into the plasma chamber and heated by a 12.6 GHz micro-wave. 
A power of a few watts is sufficient for an optimum production of the N$^\mathrm{+}$ ions. 
The ions are extracted by applying a 2 kV bias on the source and selected in mass/charge 
ratio by a dipole magnet. The selected ions are deflected by 
a 45$^o$ spherical electrostatic deflector (ED1) to be merged with the photon beam inside 
a 50 cm long interaction region (IR). Two sets of slits (S2 and S3) allow the matching of the 
size of the ion beam to the size of the photon beam. After interaction, the charge of
 the ions is analyzed by a second dipole magnet. 
 The parent ions are collected in a  Faraday cup (FC4), and the so-called photoions 
 which have gained one (or several) charge(s) 
 are counted using channel-plates.
 %
%
%
%
%

\subsection{Excitation source}
The photon beam is monochromatized synchrotron radiation from the PLEIADES beam line. 
Two undulators with 256 mm and 80 mm period deliver photons in the 10 - 100 eV and 
100 eV - 1000 eV energy ranges, respectively, with all types of polarization above 55 eV. 
The light is monochromatized by a plane grating monochromator with no entrance slit. 
High spectral purity is obtained by a combination of a quasi-periodic design for the undulators 
and the use of a varied groove depth for the plane grating. Three varied line spacing gratings 
are available for the adjustment of the resolution. An ultimate resolving power of approximately 10$^5$
is achievable at 50 eV. After monochromatization, the light is distributed and focalized to three 
different branches. On branch A, used for this experiment, the photon beam is parallel, with a 
typical spot size of 2 mm $\times$ 2 mm. The photon flux is measured using a calibrated silicon photodiode (PD). 
The photon energy is determined using a double-ionization chamber of the Samson type \cite{SamsonBook}. 
For this work, we used the 1s $\rightarrow$ $\pi$$^{*}$ transitions in the N$_{2}$ gas \cite{Sodhi1984} 
 and  $\rm 2p \rightarrow 3d$ transitions in Ar gas \cite{King1977} for calibration purposes. 
 The photon energy was corrected for Doppler shift resulting from the velocity of the N$^{+}$ ions.
The estimated accuracy of the photon energy determination is 40 meV.

\subsection{Experimental procedure}
The merged-beam set-up allows a determination of the absolute photoionization cross sections. 
At a given photon energy, the cross sections $\sigma_{\rm PI}$ are obtained from:

\begin{equation}
\sigma_{\rm PI} = \frac{S e^{2}\eta\nu q}{IJ\epsilon \int^{L}_{0} \frac{dz}{\Delta x\Delta y F(z)}}
\end{equation}

\noindent
where {\it S} is the counting rate of photoions measured with the channel-plates and 
$\eta$ is the efficiency of the photodiode. A chopper, placed at the exit of the photon beam line, 
allows to subtract from the photoions signal,  the noise produced by collisional 
ionization processes,  charge stripping on the slits or autoionizing 
decay of metastable excited states of N$^\mathrm{+}$ ions produced in the ECRIS.  Here {\it q} is the charge of 
the target ions; $\nu$ is the velocity of the ions in the interaction region determined from the 
accelerating potential applied to the ECRIS;  {\it I} is the current  produced by the photons on the calibrated photodiode.
The efficiency $\eta$ of the photodiode  was calibrated in the 10-1000 eV energy range at the 
Physikalisch-Technische Bundesanstalt (PTB) beam line at BESSY in Berlin;  {\it e} 
is the charge of the electron;  {\it J} is the current of incident ions measured in FC4; $\epsilon$ 
is the efficiency of the microchannel plates determined by comparing the counting rate 
produced by a low intensity ion beam and the current induced by the same beam in FC4; $\Delta x\Delta y F(z)$ 
is an effective beam area ({\it z} is the propagation axis of the two beams), and {\it F(z)} is a two 
dimensional form factor determined using three sets of {\it xy} scanners placed at each end and 
in the middle of the interaction region. Each scanner is a 0.2 mm width slit moved across the ion 
and the photon beams. The length {\it L} of the interaction region is fixed by applying a 200 V bias on a 
50 cm long tube placed in the interaction region, resulting in a different velocity for the photoions 
produced inside and outside the tube. {\it F(z)} is defined by:

\begin{equation}
	F_{xy} \approx \frac{\sum i_{xy} \sum j_{xy}}{\sum\sum i_{xy} j_{xy}}
\end{equation}
\noindent
where {\it i$_{xy}$}={\it i(x,y)$\Delta x$$\Delta y$} and {\it j$_{xy}$}={\it j(x,y)$\Delta x$$\Delta y$} are 
the ion and photon currents, respectively, passing through the slits and measured with FC4 for the ion 
beam and the photodiode for the photon beam, and $\Delta x$ and $\Delta y$ are the step sizes 
used to scan the slits, typically 0.2 mm. Typical values of the parameters involved in equation (1), 
measured at a  photon energy of 400 eV are given in table~\ref{Table1}.  
The contribution of N$_2^{2+}$ molecular ions in the incident beam was subtracted.
It was measured after the beam time using $^{14}$N~$^{15}$N isotopic gas to be of the order of 3 \%.

%
%
%

\begin{table}
\caption{\label{Table1} Typical values for the experimental parameters involved in evaluating 
					the absolute cross section 
                                           measured at  a photon energy of 400 eV.}
\begin{indented}
\item[]\begin{tabular}{@{}*{7}{l}}
\br
{\it S}			&30 Hz				\\
Noise		&17 Hz				\\
$\nu$		&1.6 10$^{5}$ m/s		\\
Photon flux&4.1 10$^{11}$ s$^{-1}$		\\
{\it J}&38 nA						\\
$\epsilon$&0.62					\\
{\it F}$_{xy}$&30					\\
\br
\end{tabular}
\end{indented}
\end{table}

The accuracy of the measured cross-sections is determined by statistical fluctuations on the photoion 
and noise counting rates and a systematic contribution resulting from the measurement of the 
different parameters in equation (1). The latter is estimated to be 15\% and is dominated
 by the uncertainty on the determination of the photon flux, the form factor and detector efficiency.

To record the single (double) photoionization spectra, the field in the dipole magnet  was adjusted to 
detect N$^{2+}$ ions (N$^{3+}$ ions) with the channel-plates as the photon energy was scanned. 
Two modes have been used. One with no voltage applied on the interaction tube, 
allowing a better statistic since the whole interaction length of the beams is used. 
In this mode only relative cross sections are obtained. In the second mode, the voltage 
is applied to the tube to define the interaction length $\it {L}$ (50 cm),  which allows the determination 
of the cross sections in absolute value.

\section{Theory}\label{sec:Theory}

\subsection{MCDF}\label{subsec:MCDF_Theory}
We have performed multi-configurational Dirac-Fock (MCDF) calculations 
based on full intermediate coupling in a {\it jj} basis using the code 
developed by Bruneau \cite{Bruneau1984}. 
The photoexcitation cross sections involving 1s$\rightarrow$np transitions have 
been considered for all the levels of the N$^\mathrm{+}$  ground configuration, 
namely $\rm 1s^22s^22p^2~^3P_{0,1,2}$, $\rm ^1D_2$ and $\rm ^1S_0$. 
The metastable $\rm 1s^2 2s 2p^3~^5S^o_2$ level 
also has been considered. The calculations were limited to principal quantum number up to 4. 

Only electric dipole transitions have been computed using Babushkin and Coulomb gauges, respectively. 
The oscillator strengths obtained in the two gauges differ by less than 10\%.  
Using the MCDF code we have calculated the average autoionization rate $\Gamma_{\rm av}$ (meV) 
of the $\rm 1s2s^22p^3$ configuration (with a 1s vacancy)   
to be equal to 96 meV (which corresponds to a lifetime $\tau$ of 3.43 femto-seconds (fs) 
from the uncertainty principle $\Delta\,E\, \Delta\, t \, =\, \hbar/2$).  
From the MCDF calculations synthetic spectra are 
constructed as a sum of lorentzian profiles using, for all the lines, 
the $\Gamma_{\rm av}$ value as the full width at half maximum (FWHM).
 In order to compare directly with the experimental measurements made at 300 meV and 60 meV the MCDF results 
were convoluted with gaussian functions of FWHM of 300 meV and 60 meV respectively, to simulate the measurements. A non-statistical distribution of the ground and metastable states gave
best agreement with experiment by weighting the contribution of the
 $\rm 1s^22s^22p^2~^3P$,$\rm ^1D$, $\rm ^1S$,
and $\rm 1s^22s2p^3~^5S^o$ states by (0.54, 0.11, 0.03, 0.32) respectively.  

\begin{figure}
\begin{center}
\includegraphics[scale=2.0,width=14.5cm]{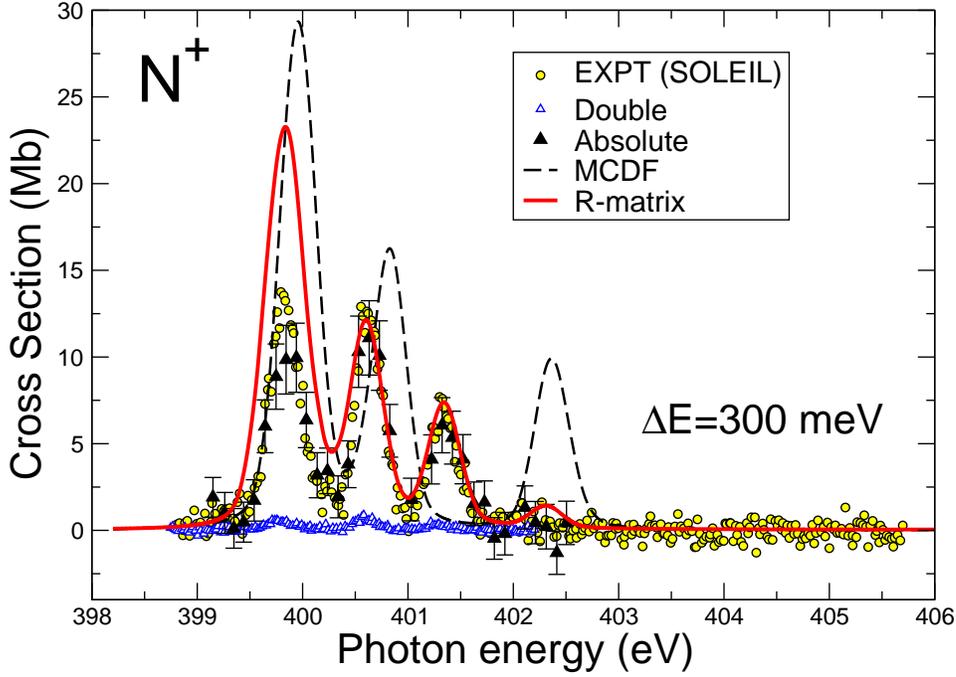}
\caption{\label{fig:300meV}(Colour online) Photoionization cross sections measured with a 300 meV band pass. 
								Open circles : single photoionization. 
								The absolute measurements (solid black triangles), single 															photoionization have been obtained with a larger energy step. 
								Open blue triangles: double photoionization. 
								The error bars give the total uncertainty of the experimental data. 
								The MCDF (dashed line) and  R-matrix  (solid line)
								calculations shown were obtained  by convolution with a Gaussian
								 profile of 300 meV FWHM and a weighting of the states 
								 (see text for details) to simulate the measurements. }

\end{center}
\end{figure}

\begin{figure}
\begin{center}
\includegraphics[scale=2.0,width=14.5cm]{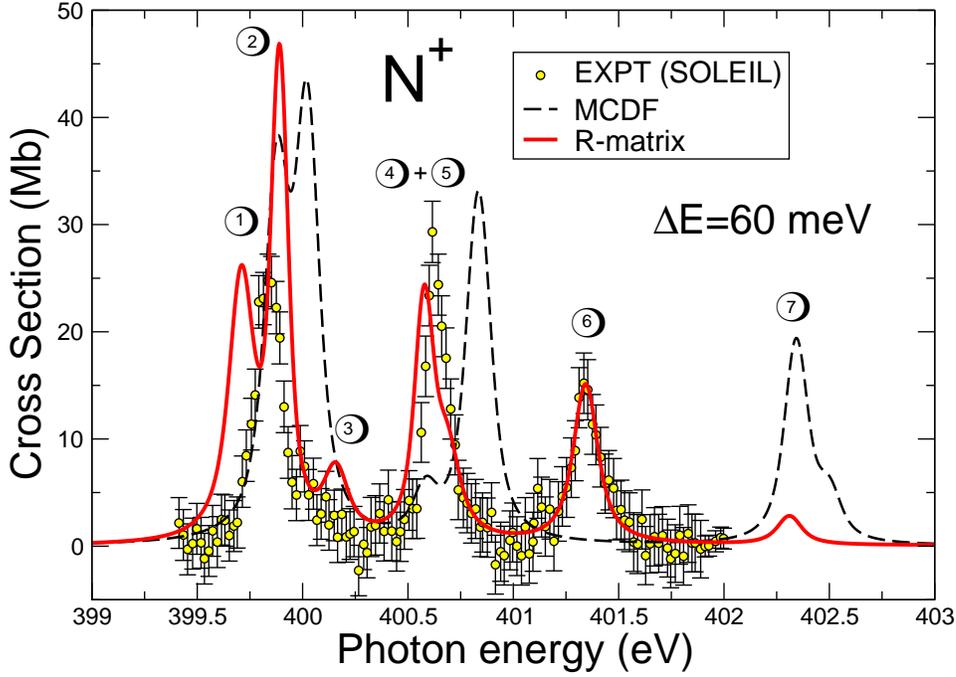}
\caption{\label{fig:60meV} (Colour online) Comparison of the single photoionization cross section (yellow circles)
								measured with 60 meV band pass (points with error 	
								bar giving the statistical uncertainty). 
								The MCDF (dashed line) and  R-matrix (solid line),
								calculations shown were obtained  by convolution with a Gaussian
								 profile of 60 meV FWMH and a weighting of the
								  states (see text for details) to simulate the measurements.
								  Table 2 gives designation of the resonances~\mycirc{1}  - \mycirc{7},}
\end{center}
\end{figure}

\subsection{R-Matrix}\label{subsec:R-Matrix_Theory}

The $R$-Matrix method \cite{rmat,codes,damp}, with an efficient parallel implementation 
of the codes \cite{ballance06} was used to determine
all the PI cross sections, for both the initial $\rm ^3P$ ground state 
and the, $\rm ^1$S, $\rm ^1$D and  $\rm ^5$S$\rm ^o$ metastable states  
in $LS$-coupling using 390-levels in the close-coupling expansion.
The Hartree-Fock $\rm 1s$, $\rm 2s$ and $\rm 2p$ tabulated orbitals of Clementi and Roetti
\cite{Clementi1974} were used together with the n=3
 orbitals of the N$^{2+}$ ion determined by energy optimization on the appropriate
hole-shell state using the atomic structure code CIV3 \cite{Hibbert1975}.  
The n=3 correlation (pseudo) orbitals are included to
account for core relaxation and additional correlation effects in the
multi-configuration interaction wavefunctions.
All the N$^\mathrm{2+}$ residual ion states were then represented
by using multi-configuration interaction wave functions. The non-relativistic
$R$-matrix approach was used to calculate the energies
of the N${^\mathrm{+}}$ bound states and the subsequent PI cross sections.
 PI cross sections  out of the $\rm 1s^22s^22p^2$\, $\rm ^3$P ground state and the  $^1$S, $^1$D 
along with the 1s$\rm ^2$2s2p$^3$ $\rm ^5$S$\rm ^o$ metastable  states
were then obtained for all total angular momentum scattering symmetries that contribute.

All our PI cross sections were determined in $LS$ - coupling 
with the parallel version \cite{ballance06}
of the R-matrix programs \cite{rmat,codes,damp} and
390 levels of the N$^\mathrm{2+}$ residual ion
included in our close-coupling calculations.  
Due to the presence of meta stable states in the beam, 
PI cross-section calculations were performed  
for both the $\rm 1s^22s^22p^2~^3P$ ground state and 
the $\rm 1s^22s^22p^2$\, $^1$S, $^1$D  
and $\rm 1s^22s2p^3$\, $\rm ^5$S$\rm ^o$ meta stable states of the 
N$^\mathrm{+}$ ion. Here again, in order to compare directly with 
the experimental measurements made at 300 meV and 60 meV the R-matriix results 
were convoluted with gaussian functions of FWHM of 300 meV 
and 60 meV respectively, to simulate the measurements. 
A non-statistical distribution of the ground and metastable states gave
best agreement with experiment by weighting the contribution of the
 $\rm 1s^22s^22p^2~^3P$,$\rm ^1D$, $\rm ^1S$,
and $\rm 1s^22s2p^3~^5S^o$ states by (0.54, 0.11, 0.03, 0.32) respectively.  

The scattering wavefunctions were generated by
allowing all possible two-electron promotions out of select base
configurations of N$^\mathrm{+}$ into the orbital set employed.
Scattering calculations were performed with twenty
continuum functions and a boundary radius of 6.2 Bohr radii.
For both the $\rm ^3$P ground state and the  $\rm ^1$S, $\rm ^1$D,  $\rm ^5$S$\rm ^o$ 
metastable states the outer region electron-ion collision
problem was solved (in the resonance region below and
 between all the thresholds) using a suitably chosen fine
energy mesh of 2$\times$10$^{-7}$ Rydbergs ($\approx$ 2.72 $\mu$eV)
to fully resolve all the resonance structure in the PI cross sections.  
Radiation and Auger damping were also
included in the cross section calculations.  
Finally the total spectrum was obtained assuming a  weighting  of the
 $\rm 1s^22s^22p^2~^3P$,$\rm ^1D$, $\rm ^1S$,
and $\rm 1s^22s2p^3~^5S^o$ states by (0.54, 0.11, 0.03, 0.32) respectively in order to compare
directly with the PI cross section measurements obtained  using the ion beam 
produced from the ECRIS at the SOLEIL radiation facility.

The multi-channel R-matrix QB technique (applicable to atomic and molecular complexes)
of Berrington and co-workers \cite{keith1996,keith1998,keith1999}
was used to determine the resonance parameters. The resonance width $\Gamma$
may be determined from the inverse of the energy derivative of the eigenphase
sum $\delta$ at the position of the resonance energy $E_r$ via
\begin{equation}
\Gamma = 2\left[{\frac{d\delta}{dE}}\right]^{-1}_{E=E_r} = 2 [\delta^{\prime}]^{-1}_{E=E_r} \quad.
\end{equation}

The results for the resonance parameters  determined from the QB method  are presented in Table 2.

\section{Results and Discussion}\label{sec:Results}
In the entire 388-430 eV photon energy range we explored, only three intense structures were 
observed around 400 eV. Figure 2 shows our experimental and theoretical (MCDF and R-matrix) results over the photon 
energy range 398 eV -- 406 eV.  The experimental results were taken with a photon energy resolution of 300 meV. 
 The black triangles show the measurements of the single PI cross section obtained in the absolute mode. 
 The yellow circles have been obtained in the relative mode with a smaller energy step. 
 They have been put on the absolute scale keeping the same area under the absolute and relative measurements. 
 The open blue triangles give the double PI cross section determined in the relative mode. 
 To place them on the absolute scale it was assumed that the length of the interaction region 
 was the same while recording the relative single and double PI cross sections. 

 In order to compare directly with the experimental measurements both our MCDF and R-matrix results 
were convoluted with a gaussian function of FWHM of 300 meV to simulate the measurements. 
Furthermore, we found that a non-statistical distribution of the ground and metastable states gave
best agreement with experiment by weighting the contribution of the
 $\rm 1s^22s^22p^2~^3P$,$\rm ^1D$, $\rm ^1S$,
and $\rm 1s^22s2p^3~^5S^o$ states by (0.54, 0.11, 0.03, 0.32) respectively.
A statistical weighting showed greater disparity between theory and experiment.
Figure 3 shows the same experimental and theoretical 
(MCDF and R-matrix) single PI cross sections results over the photon 
energy range 399 eV -- 403 eV but at the higher resolution of 60 meV. 
Voigt profiles were used to fit the peaks in the experimental data, to extract the linewidths,
assuming a Gaussian instrumental contribution of 60 meV for each peak.
The experimental data, obtained in the relative mode, were normalised keeping the same total area 
under the experimental single PI cross section measured with 60 meV and 300 meV band 
pass (yellow circles on figures 2 and 3). The error bars give the statistical uncertainty. 
The measured double PI cross section is only 4\% of the single PI, these data are shown in figure 2 by the open blue triangles.
 Here again theory has been convoluted with a gaussian of 60 meV FWHM and the weighting of the states used as before.
Both MCDF and R-matrix calculations are in good qualitative agreement with the experimental results. 
The position of the lines is better described by the R-matrix calculations, from matching the calculated 
and experimental ionization thresholds. The R-matrix calculations include Auger and radiation damping, 
missing from the MCDF calculations and give lines intensity in closer agreement with experiment. 
The R-matrix calculations indicate radiation damping can contribute up to 30\%.
%
%
%

\begin{table}
\caption{\label{tab:fit2} Resonance energies $E_{\rm ph}^{\rm (res)}$ (eV),
         and the natural linewidths $\Gamma$ (meV) for the core photoexcited n=2 states  
         of N$^{+}$ ions in the photon energy region 399 eV to  403 eV. 
         The experimental uncertainty is $\pm$ 40 meV for the resonance energies.  
         The earlier MCDF calculations of Chen and co-workers \cite{Chen1997}  
         are included for comparison purposes.}
 \lineup
  \begin{tabular}{ccr@{\,}c@{\,}llcl}
\br
 Resonance    & & \multicolumn{3}{c}{SOLEIL}   & \multicolumn{1}{c}{R-matrix} 		& \multicolumn{2}{c}{MCDF}\\
 (Label)      & & \multicolumn{3}{c}{(Experiment$^{\dagger}$)}      & \multicolumn{1}{c}{(Theory)} 		& \multicolumn{2}{c}{(Theory)}\\
 \ns
 \mr
  $\rm 1s^22s^22p^2\, ^3$P \,$\rightarrow$\, $\rm 1s2s^22p^3\, ^3D^{\rm o}$				& $E_{\rm ph}^{\rm (res)}$         
             & 		& --				& 	& 399.706$^{a}$    		& 	& 399.894$^{b}$   \\
  ~\mycirc{1} &           &   					&  						&	&     					&	& 394.665$^{c}$		      \\
 & $\Gamma$
                  & \;\0\0\    & --    & 		     					& \0\0124$^{a}$  		 & 	&\078$^{b}$    \\
 	    &           &   					&  						&	&     		&	& 207$^{c}$		      \\
  $\rm 1s^22s2p^3\,^5S^{\rm o}$ \,$\rightarrow$\, $\rm 1s2s2p^4\, ^5P$				& $E_{\rm ph}^{\rm (res)}$         
              & 		& --								&  	& 399.891$^{a}$    		& 	& 400.834 $^{b}$   \\
 ~\mycirc{2} 	    &           &   					&  						&	&     		&	& 401.568 $^{c}$		      \\
 & $\Gamma$
                  & \;\0\0\0    & --    &  		     					& \0\062$^{a}$  		 & 	& 104$^{b}$    \\
 	    &           &   					&  						&	&     		 &	& \043$^{c}$		      \\              
   Average				& $E_{\rm ph}^{\rm (res)}$         
                      & 	 		& 399.844 $\pm$ 0.04$^{\dagger}$		& 	& 399.799$^{d}$    		& 	& 400.364$^{e}$   \\
  ~\mycirc{1} + \mycirc{2} &           &   					&  						&	&     					&	&   \\
 & $\Gamma$
                  & \;\0\0\    & 138  $\pm$  \041$^{\dagger}$    & 		     					& \0\093$^{d}$  		 & 	&\091$^{e}$    \\
 \\
 $\rm 1s^22s^22p^2\,^1$S \,$\rightarrow$\, $\rm 1s2s^22p^3\, ^1P^{\rm o}$				& $E_{\rm ph}^{\rm (res)}$         
              	 & 		& --								&  	& 400.159$^{a}$    		& 	& 400.583$^{b}$   \\
 ~\mycirc{3}	    &           &   					&  						&	&     		&	& 400.146$^{c}$		      \\
 & $\Gamma$
                  & \;\0\0\0    & --    &  		     					& \0\0115$^{a}$  		 & 	& 106$^{b}$    \\
 	    &           &   					&  						&	&     		&	& 164 $^{c}$		      \\
  $\rm 1s^22s^22p^2\, ^3$P \,$\rightarrow$\, $\rm 1s2s^22p^3\, ^3S^{\rm o}$				& $E_{\rm ph}^{\rm (res)}$         
             & 		& --								&  	& 400.579$^{a}$    		& 	& 399.999$^{b}$   \\
  ~\mycirc{4} 	    &           &   					&  						&	&     		&	& 396.13 $^{c}$		      \\
 & $\Gamma$
                  & \;\0\0\0    & --    &  		     					& \0\078$^{a}$  		 & 	&\064$^{b}$    \\
 	    &           &   					&  						&	&     		&	& 122$^{c}$		      \\
 $\rm 1s^22s^22p^2\,^1$D \,$\rightarrow$\, $\rm 1s2s^22p^3\, ^1D^{\rm o}$				& $E_{\rm ph}^{\rm (res)}$         
              & 		& 	--		&  	& 400.681$^{a}$    		& 	& 400.031$^{b}$   \\
 ~\mycirc{5} 	    &           &   					&  						&	&     			&	& 397.859$^{c}$		      \\
 & $\Gamma$
                  & \;\0\0\0    & --   &  		     					& \0\0105$^{a}$  		 & 	& 108$^{b}$    \\
 	    &           &   					&  						&	&     		 &	&170$^{c}$		      \\
   Average				& $E_{\rm ph}^{\rm (res)}$         
                      & 	 		& 400.633	$\pm$ 0.04$^{\dagger}$		& 	& 400.630$^{d}$    		& 	& 400.015$^{e}$   \\
  ~\mycirc{4} + \mycirc{5} &           &   					&  						&	&     					&	&   \\
 & $\Gamma$
                  & \;\0\0\    &\0 86 $\pm$ \010$^{\dagger}$   & 		     					& \0\092$^{d}$  		 & 	&86$^{e}$    \\
 \\
   $\rm 1s^22s^22p^2\, ^3$P \,$\rightarrow$\, $\rm 1s2s^22p^3\, ^3P^{\rm o}$				& $E_{\rm ph}^{\rm (res)}$         
             & 		& 401.346 $\pm$ 0.04$^{\dagger}$			&  	& 401.347$^{a}$    		& 	& 402.342$^{b}$   \\
  ~\mycirc{6} 	    &           &   					&  						&	&     				&	& 396.430$^{c}$		      \\
 & $\Gamma$
                  & \;\0\0\0    & 143 $\pm$ 21$^{\dagger}$    &  		     			& \0\0121$^{a}$  		 & 	& 110$^{b}$    \\
 	    &           &   					&  						&	&     					&	& 200$^{c}$		      \\
\\	    
   $\rm 1s^22s^22p^2\,^1$D \,$\rightarrow$\, $\rm 1s2s^22p^3\, ^1P^{\rm o}$				& $E_{\rm ph}^{\rm (res)}$         
             & 		& --								&  	& 402.320$^{a}$    		& 	& 402.503$^{b}$   \\
  ~\mycirc{7} 	    &           &   					&  						&	&     		&	& 399.619$^{c}$		      \\
\\
 & $\Gamma$
                  & \;\0\0\0    & --    &  		     					& \0\0132$^{a}$  		 & 	& 107$^{b}$    \\
 	    &           &   					&  						&	&     		&	& 164$^{c}$		      \\
\br
\end{tabular}
~\\
$^{\dagger}$SOLEIL, experimental lines are not pure, see text for discussion.\\
$^{a}$R-matrix  $LS$-coupling,  present work.\\
$^{b}$MCDF, present work.\\
$^{c}$MCDF, Chen and co-workers. \cite{Chen1997}\\
$^{d}$ Average R-matrix, present work\\
$^{e}$ Average MCDF, present work\\
\end{table}

In Table 2  we present the MCDF theoretical values for the resonance energies and linewidths 
from our work  which were obtained by fitting the various peaks found in the cross sections to
 Fano profiles for overlapping resonances. This enabled the extraction of the 
 resonance energies and natural linewidths for the various states.
The assignment of the peaks in the PI cross sections (due to 1s $\rightarrow$ 2p 
photo-excitation) is based on our R-matrix theoretical work, since it was not possible to 
extract experimental values for each individual state as the lines are a blend of different states.
The first strong peak at about 399.85 eV  is primarily a blend of the 
resonance $\mycirc{1}$,   $\rm 1s2s^22p^3~^3D^o$, resulting from photo-excitation of the ground $\rm 1s^22s^22p^2~^3P$ state
along with that from resonance $\mycirc{2}$, $\rm 1s2s2p^4~^5P$ from the $\rm 1s^22s2p^3~^5S^o$ metastable state.
This is clearly seen in the higher resolution experimental data taken at 60 meV (figure 3) 
as opposed to the measurements made at the lower resolution of 300 meV (figure 2).
The valley between the first and second peak is filled with the $\rm 1s2s^22p^3~^1P^o$
resonance $\mycirc{3}$ resulting from photo-excitation of the $\rm 1s^22s^22p^2~^1S$ metastable.  
The next strong peak at about 400.63 eV is a blend
of the $\rm 1s2s^22p^3~^3S^o$ resonance $\mycirc{4}$, resulting  from photo-excitation from the ground  $\rm 1s^22s^22p^2~^3P$ 
and the $\rm 1s2s^22p^3~^1D^o$  resonance $\mycirc{5}$, caused by photo-excitation of the $\rm 1s^22s^22p^2~^1D$ 
metastable state. The third peak at about 401.35 eV is due to the resonance $\mycirc{6}$, $\rm 1s2s^22p^3~^3P^o$, resulting 
from photo-excitation of the $\rm 1s^22s^22p^2~^3P$ ground state.  Finally a broad  resonance peak $\mycirc{7}$, at 
about 402.3 eV (barely visible in the experimental data, figure 2), resulting from 
the photo-excitation of the $\rm 1s^22s^22p^2~^1D$ metastable
is  assigned to the $\rm 1s2s^22p^3~^1P^o$ resonance state.
 These assignments are summarized in the first column of Table 2. 

Experimental linewidths were extracted from the ion-yield measurements in figure 3
by fitting Voigt profiles to the peaks assuming a Gaussian instrumental contribution of 60 meV for each peak.
The first  resonance peak in the experimental data at 399.844 $\pm$ 0.04 eV has a linewidth of 138  $\pm$  41 meV, 
the second peak  located at 400.633 $\pm$ 0.04 eV has a linewidth 86 $\pm$ 10 meV 
and the third peak at 401.346 $\pm$ 0.04 has a linewidth of 143 $\pm$ 21meV.
The linewidth gives only an upper limit of the natural width since several lines can contribute to the structures. 
According to the R-matrix calculations, only peak 3 in the experimental data is pure, 
and following MCDF calculations, only peak 2, which may be due to the S $\rightarrow$ P dipole selection rules.
Table 2 presents the theoretical results for the resonance positions and widths 
from our present investigations.  For comparison purposes we have also 
included the results from the earlier MCDF calculations of Chen and co-workers \cite{Chen1997}.
From the R-matrix results we see that the average autoionization rate for the $\rm 1s2s^22p^3$ 
configuration (with a 1s vacancy)  is 112 meV compared to a value of 96 meV determined 
from our present MCDF  calculations.

In Table 2 the results from  our present theoretical predictions 
from both the MCDF and R-matrix methods overall indicate 
they are in suitable agreement with each other. Futhermore, It is seen that the R-matrix results give
closer agreement with experiment for the resonance energies of all the peaks, deviating at most by 
45 meV (c.f. the avearge for$ \mycirc{1}$ +$ \mycirc{2}$), which is just 
outside the experimental error of 40 meV.  All the other resonance energies from the 
R-matrix work are within the 40 meV experimental uncertainty however greater disparity is seen between 
experiment and the MCDF results.  The use of a much larger basis in the R-matrix work
along with coupling to the continuum could possibly improve the agreement with 
experiment but vastly increases the  computational complexity of the problem.
As seen from Table 2 discrepancies occur in the position and autoionization linewidth of the resonances from our  work
when compared to the early MCDF calculations of Chen et al. \cite{Chen1997}.  This was also highlighted by 
Garcia and co-workers \cite{Witthoeft2009} where outstanding discrepancies
were found with that MCDF work, in particular wavelengths
and A-coefficients for this C-like ion, which those authors believed to be
due to numerical error by Chen et al. \cite{Chen1997}.

\section{Conclusions}\label{sec:Conclusions}
Photoionization of C-like nitrogen ions, N$^{+}$, has been investigated 
using state-of-the-art experimental and theoretical methods.
High-resolution spectroscopy was performed with E/$\Delta$E = 7000 at the SOLEIL synchrotron 
radiation facility, in Saint-Aubin, France, covering the
energy ranges 388 eV to 430 eV where several strong 
peaks in the cross sections were found in the energy region around 400 eV.
 For these observed peaks, suitable agreement is found between the present theoretical and
experimental results both on the photon-energy scale and on the
 PI cross-section scale for this prototype C-like system.
The strength of the present study is in its excellent experimental resolving power coupled
with theoretical predictions using the MCDF and R-matrix method.  Given that the present results have been 
benchmarked with high resolution experimental data and with state-of-the-art theoretical methods 
they would be suitable to be included in astrophysical modelling codes such as CLOUDY
 \cite{Ferland1998,Ferland2003} and XSTAR \cite{Kallman2001}. The theoretical R-matrix 
 cross sections for the individual states are available by contacting one of 
 the authors, B M McLaughlin, b.mclaughlin$@$qub.ac.uk.

\ack
Data collection was performed on the PLEIADES beamline, at the SOLEIL Synchrotron 
radiation facility in Saint-Aubin, France.  The authors would like to thank the staff of SOLEIL and, 
in particular, the staff of the PLEIADES beam line for their helpful assistance.
The experimental research work described here has been supported 
by Triangle de le Physique contract 2007-001T.
B M McLaughlin acknowledges support by the US
National Science Foundation through a grant to ITAMP
at the Harvard-Smithsonian Center for Astrophysics.
The computational work was carried out at the National Energy Research Scientific
Computing Center in Oakland, CA, USA and on the Tera-grid at
the National Institute for Computational Sciences (NICS) in Knoxville, TN, USA,
which is supported in part by the US National Science Foundation.

%
%
%
%

\bibliographystyle{iopart-num}

\bibliography{nplus}
\end{document}